\documentclass[11pt]{amsart}

\usepackage[margin=1in]{geometry}

\usepackage{amsmath,comment}
\usepackage{amssymb}
\usepackage{amsthm}
\usepackage{tikz}
\usepackage{pgfplots}
\usepackage{graphicx}
\usepackage{caption}
\usepackage{subcaption}
\usepackage{enumerate}
\usepackage{mhchem}
\usepackage{framed}

\usepackage{hyperref}

\newcommand{\reals}{\mathbb{R}}

\newcommand{\ra}{\rightarrow}

\newcommand{\etal}{{\em et al.}}

\usepackage[useprefix=true,style=numeric-comp]{biblatex}
\usepackage{float}
\usepackage{wrapfig}
\usetikzlibrary{positioning}

\newcommand\img{\mathrm{im}}
\newcommand\dfn\textit

\newcommand\from\leftarrow

\graphicspath{{./../}}

\makeatletter
\g@addto@macro\maketitle{\thispagestyle{empty}\setcounter{page}{0}}
\makeatother

\begin{document}

\title[Tagging Regimes in Dynamical Systems]{Automatic recognition and tagging of topologically different regimes in dynamical systems}


\author{Jesse J. Berwald$^*$}
\address{Institute for Mathematics and its Applications, University of Minnesota\\Minneapolis, Minnesota}
\thanks{$^*$Research of JJB partially supported by the Mathematics and Climate Research Network under grant NSF DMS-0940363}
\email{jberwald@ima.umn.edu}
%
\author{Marian Gidea$^\dag$}
\address{Yeshiva University\\New York City, New York}
\thanks{$^\dag$Research of MG was partially supported by NSF grants: DMS-1201357 and DMS-0940363}\email{Marian.Gidea@yu.edu}
\author{Mikael Vejdemo-Johansson$^\diamondsuit$}
\address{AI Laboratory, Jo{\v z}ef Stefan Institute, Ljubljana, Slovenia\\
  Computer Vision and Active Perception Laboratory \\ KTH Royal Institute of Technology, Stockholm, Sweden}
\thanks{$^\diamondsuit$Research of MVJ supported by Toposys grant FP7-ICT-318493-STREP}
\email{mvj@kth.se}

\begin{abstract}
Complex systems are commonly modeled using nonlinear dynamical
systems. These models are often high-dimensional and chaotic.  An
important goal in studying physical systems through the lens of
mathematical models is to determine when the system undergoes
changes in qualitative behavior.  A detailed description of the
dynamics can be difficult or impossible to obtain for high-dimensional
and chaotic systems. Therefore, a more sensible goal is to recognize
and mark transitions of a system between qualitatively different 
regimes of behavior. In practice, one is interested in developing techniques for
detection of such transitions from sparse observations, possibly
contaminated by noise. In this paper we develop a framework to
accurately tag different regimes of complex systems based on
topological features.  In particular, our framework works with a high
degree of success in picking out a cyclically orbiting regime from a
stationary equilibrium regime in high-dimensional stochastic dynamical
systems.


\end{abstract}

\maketitle

\newpage

\section{Introduction}
\label{sec:introduction}

Critical transitions are abrupt changes in the behavior of nonlinear
systems that arise after small changes in the parameters of a
system. They are natural phenomenon which occur in across a vast range
of spatial and temporal scales. Natural systems that exhibit sudden
shifts in their behavior include the Earth's climate, changes in ocean
currents, large and sudden shifts in plant and animal populations, and
the domino-like collapses observed in financial
markets~\cite{Scheffer2009}. In a climatic example, data indicate
that the Earth's climate has swung between a ``snowball Earth'' and a
``tropical'' Earth numerous times in its history on a geologic time
scale that is considered rapid~\cite{Hoffman1998}. Also on a global
scale, an abrupt change in the strength and direction of the Gulf
Stream as a result of climate change would prove catastrophic for the
European climate~\cite{Vellinga2002}. Evidence suggests that such a change was
partially responsible for the three-hundred year Little Ice Age in
Europe beginning in the 17th century~\cite{Keigwin1996}. Ecology provides another
good source of examples of catastrophic change; for instance,
eutrophication of a lake occurs when nutrient-rich pollution reaches a
critical threshold, at which point water clarity is suddenly and
greatly reduced due to a bloom of algae, which in turn kills submerged
flora~\cite{Scheffer2009,Scheffer2001b}.

The aim of this paper is to develop robust methods by which to
characterize and detect critical transitions between two or more
regimes in the evolution of a dynamical systems. Formally, critical
transitions as seen in natural system are associated to {\em
  bifurcations} in dynamical systems models. Bifucations occur when
basins of attraction collide due to a small change in parameters,
which can lead to the disappearance of one stable region and cause the
stability of the system to undergo a sudden, or critical,
transition. In models with stochasticity, bifurcations are
particularly difficult to define, since the change in regime will
depend on the particular realization of the underlying random
variable. Some current approaches~\cite{ChenFu_randomDS} use
topological methods to devise sufficient conditions for bifurcations
in stochastic dynamical systems, by using all possible realizations of
the random variable. From a practical point of view, when the data
generated by a system is acquired from a single, or a very limited
number of realizations, such an approach may not be suitable. An
additional challenge encountered in many real world systems is
sparsity of, and noise pollution in, the available data. In this paper
we outline a novel method to detect critical transitions in dynamical
systems with additive noise or time series measured from real-world
sources. Our approach is based on combining the theory of persistence
homology with machine learning techniques.

The importance of detection and prediction of critical transitions
from observational data in the context of ecology and climate science
has been emphasized in a series of recent papers by Dakos, Ditlevsen,
Scheffer
~\cite{Dakos2008,Ditlevsen2010,Lenton2008,Livina2007,Scheffer2001b,Scheffer2009}.
The main results of Scheffer, Dakos, and collaborators concern
1-dimensional time series in which a sequence of sliding windows is
used to study changes in the statistical properties of the system over
time. They demonstrate a correlation between the resilience of the
systems under study and changes in the variance and autocorrelation
measured across windows, especially in when the system is in close
proximity to a bifurcation. A challenging aspect of their approach is
the a priori lack of robustness of the statical methods involved since
they rely on a number of choices of window size, detrending method,
and visual interpretation of the results.

In this paper, we apply techniques from topological data analysis
(TDA) to study, first, bifurcations in two classical dynamical systems
with additive noise; and second, measurements of real-world,
high-dimensional climate phenomena for which a critical transition
manifests in the data. The main contribution of this work is the
development of a methodology independent of dimension to detect the
presence of a critical transitions.  By studying the {\em persistent
  homology} of the point cloud data over windows (subsets) of a time
series we can analyze and detect topologically distinct regimes of
the behavior of the dynamical system. Another important feature of our
approach is that it is robust, in the sense that data sets that are
very close to one another yield topological objects that are also very
close to one another, relative to some appropriate metric~\cite{Cohen-Steiner2007}.

In the sections that follow, the paper is broken into two main
parts. In Section~\ref{sec:background} we summarize the relevant
dynamical systems background as well as persistent homology, which
allows the encoding of topological information in the form of {\em  persistence diagrams}. In Section~\ref{sec:learn-topol-diff} we
describe the basis for our classifier in the context of machine
learning and the selection of relevant features from persistence
diagrams, their relation to the underlying dynamical system, and the
use of machine learning to classify a given system using the selected
features.  
We focus on the case of periodic and quasi-periodic
phenomena, and use degree 1 Betti numbers for detection of critical transitions. In
Sections~\ref{sec:results}, we demonstrate the
effectiveness of our algorithms in detecting bifurcations in noisy
systems with conceptual computational models and follow this with
analysis of real-world time series data. We conclude with a discussion
of results and future directions in Section~\ref{sec:conclusion}.

\section{Background}
\label{sec:background}

\subsection{Dynamical systems}
\label{sec:ds}

We recall some basic facts about parameter-dependent differential equation, both deterministic and stochastic. Simple physical systems are often described by ordinary differential equations that depend explicitly on one or more parameters,  of the form
\begin{align}\label{eq:ode-param}
  \dot{x} = f( x,\lambda),
\end{align}
where $f : \reals^n\times\Lambda \ra \reals^n$ is a $C^1$-function in
all variables and the parameter space $\Lambda$ is a subset of $\reals^m$.
The corresponding flow, denoted
$\varphi=\varphi(t,x;\lambda)$ depends in $C^1$-fashion on time,
initial condition and parameter, and satisfies
\begin{itemize}
\item[(i)] $\varphi(0;\lambda)=\textrm{id}_{\reals^n}$;
\item[(ii)] $\varphi(t+s;\lambda)=\varphi(t;\lambda)\circ\varphi(s;\lambda)$, for all $s,t\in \reals$,
\end{itemize}
where we denote by $\varphi(t;\lambda)$ the diffeomorphism $x\in \reals^n\mapsto \varphi(t,x;\lambda)\in \reals^n$.

Two distinct flows $\varphi$ and
$\psi$ are called  {\em topologically conjugate} if there exists a homeomorphism $h: \reals^n\to
\reals^n$ such that $h\circ \phi_t=\psi_t\circ h$ for all $x$. The flows $\varphi$ and
$\psi$ are {\em topologically equivalent}  if there exists a homeomorphism $h: \reals^n\to
\reals^n$ and a continuous time-rescaling function $\tau(x,t)$ which is strictly increasing in $t$   such that
$h\circ \varphi_{t}(x)=\psi_{\tau(x,t)}\circ h(x)$ for all $(t,x)$. If the two flows are topologically equivalent,
then $h$ maps the orbits of one flow onto the orbits of the other flow, preserving direction of time but not necessarily parametrization by time.

As the parameter $\lambda$ of \eqref{eq:ode-param} varies, the topological equivalence between the flows $\varphi(\cdot;\lambda)$ for different values of $\lambda$ may cease to exist.
When
this happens, we say that a {\em bifurcation} has occurred. The value
$\lambda^*$ of the parameter that marks the change of topology under variation of
 parameters is referred to as a {\em bifurcation value}.
More precisely, $\lambda^*$ is a bifurcation value if for any neighborhood $V$ of $\lambda^*$
there is a parameter value $\lambda\in V\setminus \{\lambda^*\}$ such that $\phi(\cdot;\lambda)$ and
$\phi(\cdot;\lambda^*)$ are not topologically equivalent.

Some bifurcations can be detected by analyzing the behavior of the flow in
a small neighborhood of an equilibrium point; these are referred to
as local bifurcations. Others require analyzing the whole phase
 portrait; those are referred to as global bifurcations.  An example
 of a local bifurcation is the Hopf bifurcation, when a
 stable (unstable) equilibrium point becomes unstable (stable) and a
 stable (unstable) periodic orbit is born for some value of the
 parameter. An example of a global bifurcation is when a connecting
 orbit between two equilibrium points of saddle type breaks down for
 some value of the parameter. Global bifurcations can also more complicated sets, such as attractors,
 which can appear, disappear, merge into, or split from, one another.
To recall, an attractor is a set $A\subseteq \reals^n$ invariant under the flow, for which there is a neighborhood $U$ of $A$ such that $\phi_t(x)\to A$ as $t\to\infty$ for all $x\in U$.

Physical systems that are perturbed by external noise can be modeled
by stochastic differential equations (SDE) which may also depend on
parameters. Below, we detail the ways in which bifurcations in SDEs
mirror those in ODEs. We consider the simplest type of SDE's with
additive Gaussian (white) noise
\begin{align}\label{eq:gen-sys}
\dot{x} = f(x;\lambda) + \sigma \eta_t,
\end{align}
where $\eta_t$ is Gaussian noise with mean 0 and
standard deviation 1, and $\sigma$ is the noise intensity of the equation. This can be written as
\begin{align}\label{eq:gen-sys2}
dx = f(x;\lambda) dt + \sigma dW_t,
\end{align}
where $W_t$ is a standard Brownian motion (that is, $dW_t=\eta_tdt$).
If $f$ is uniformly Lipschitz continuous, the equation \eqref{eq:gen-sys} with
initial condition $x(t_0)=x_0$ has a solution $x(t)$ that depends
continuously on time, initial condition, and parameter. The dependence on time is only H\"older continuous.
Moreover, the solution
depends on the realization of the underlying Brownian motion. Another
remark is that when $\sigma\to 0$ the solution of the initial
value problem for an SDE approaches uniformly the corresponding solution
of the ODE (see \cite{Arnold,Evans}).

Let $\Omega=C(\reals,\reals)$ be the space of continuous functions on
reals, regarded as the path space of Brownian motions $\eta_t$,
equipped with the Wiener measure. More details can be found in
Evans~\cite{Evans}. The probabilistic nature of SDEs forces a change
in how one defines bifurcations. Briefly, the solutions of the stochastic differential equation
\eqref{eq:gen-sys} define what is known as a {\em cocycle} $\varphi:\reals\times
\Omega\times\reals^n\times\Lambda\to\reals^n$ characterized by the
following two properties:
\begin{itemize}
\item[(i)] $\varphi(0,\omega;\lambda)=\textrm{id}_{\reals^n}$ for all $\omega\in \Omega$;
\item[(ii)] $\varphi(t+s,\omega;\lambda)=\varphi(t,\theta_t\omega;\lambda)\circ\varphi(s,\omega;\lambda)$, for all $s,t\in \reals$ and all $\omega\in \Omega$,
\end{itemize}
where $(\theta_t\omega)(s)=\omega(s+t)-\omega(s)$ for $\omega\in\Omega$. Here $\varphi(0,\omega;\lambda)$ is the mapping $x\in \reals^n\mapsto \varphi(0,\omega,x;\lambda)\in \reals^n$.
When $\Omega$ reduces to a point, the cocycle property coincides with the usual flow property.

Two cocycles $\varphi$ and $\psi$ are said to be topologically conjugate if there exists a family $\{h(\omega):\reals^n\to\reals^n\}_{\omega\in\Omega}$ of homeomorphisms such that the mappings $\omega\mapsto h(\omega)(x_1)$ and   $\omega\mapsto h^{-1}(\omega)(x_2)$ are measurable for all $x_1,x_2\in\reals^n$, and the cocycles
$\varphi$  and $\psi$ are cohomologous, i.e.,
$\psi(t,\omega,h(\omega)(x))=h(t,\theta_t\omega,\varphi(t,\omega,x))$ for all $x\in \reals^n$ and almost all $\omega\in\Omega$.  A bifurcation value  for \eqref{eq:gen-sys} is a value $\lambda^*$ such that for any neighborhood $V$ of $\lambda^*$ there is $\lambda\in V\setminus\{\lambda^*\}$ such that $\varphi(\cdot;\lambda)$ is not topologically conjugate with $\varphi(\cdot;\lambda^*)$.
We notice that, unlike for flows, topological conjugacy of cocycles does not involve a time reparametrization.
This is because in the stochastic case periodic orbits exist with zero probability. Additionally, we remark that the stochastic version of topological conjugacy requires an entire family of homeomorphisms to satisfy the measurability condition.

Crucially, in practice this definition is difficult to verify. One reason is that, when dealing with experimental data, it is not possible to generate a significantly large number of realizations of an experiment, but only a very limited number of them. In climate data, for example, only very few sets of proxy temperature measurements are available. In the subsequent sections, we will use topological tools to devise a practical method to characterize bifurcations in parameter-dependent SDEs.
Instead of  focusing on a single bifurcation value, we will examine a range of parameters  and we will assess whether there is a significant change in the topological features of phase space over that range. In particular, we will look at attractors corresponding to different values of the parameter within the range, and we will characterize their topology in terms of the homology groups.
Moreover, instead of looking at continuous solutions of ODEs, we will look at their time discretizations, so instead of paths, we will examine discrete sets of points.

\subsection{Change of topology of attractors undergoing bifurcations}
We consider the following  specific situation. We start with the system \eqref{eq:ode-param} where the parameter space $\Lambda$ is some interval in $\reals$.
We assume that for some $\lambda =\lambda^*\in \Lambda$ the system   undergoes a bifurcation, and for $\lambda\in\Lambda\setminus\{\lambda^*\}$  the system has an attractor $A_\lambda$ that depends on $\lambda$.
Moreover, we assume that there is a change in the topology of the attractor, as follows. By $H_*$ we mean the singular homology of a topological space.
Suppose that, for some $\delta>0$ we have:
 \begin{enumerate}[(i)]
 \item For $ \lambda < \lambda^*-\delta$, the system
   \eqref{eq:ode-param} has an attractor $A_{\lambda}$, with
   $H_*(A_{\lambda})$ constant;
 \item For $\lambda^*+\delta< \lambda' $, the system
   \eqref{eq:ode-param} has an attractor $A_{\lambda'}$, with
   $H_*(A_{\lambda'})$ constant;
 \item For each $\hat\lambda \in (\lambda^* - \delta,
   \lambda^*+\delta)\setminus \{\lambda^*\}$ there exists $k$ such
   that $H_*(A_{\hat{\lambda}}) \ne H_*(A_{\lambda})$ or
   $H_*(A_{\hat{\lambda}}) \ne H_*(A_{\lambda'})$.
  \end{enumerate}

In the above, in (iii), whether $H_*(A_{\lambda}) =H_*(A_{\lambda'})$ depends on
the system. For a one dimensional system displaying hysteresis, the
topology of the two attractors is indeed identical. On the other hand, in the
case of a Hopf bifurcation, as in Section~\ref{sec:hopf}, the stable
equilibrium point changes to a stable limit cycle meaning
$H_*(A_{\lambda}) \ne H_*(A_{\lambda'})$. The crucial observation is
that near a bifurcation  the homology of an attractor of the system changes due to the
inherent instabilities of the system.
However, computing the homology of an attractor is difficult.
Instead, we will measure the change of topology of attractors via persistent homology, as explained below.
Moreover, we will consider not only deterministic systems as described by \eqref{eq:ode-param}
but also random dynamical systems, as described by \eqref{eq:gen-sys}.

\subsection{From dynamical systems to point cloud data}\label{point-clouds}

We consider an ODE given by \eqref{eq:ode-param}, or an  SDE, given by \eqref{eq:gen-sys}, with $\sigma$
sufficiently small. We consider the time evolution of a system, starting with some initial condition, and with the parameter $\lambda$ slowly evolving. To the system output
 $(t,\varphi_t(x))$, we apply a time discretization
$t_0,t_1,\ldots, t_N$ with equal time increments $\Delta t$ that are
sufficiently small. Thus, instead of a trajectory of the flow associated to
\eqref{eq:gen-sys}, we consider an orbit $x_0, x_1,\ldots, x_N$  of the $(\Delta t)$-time map $F$,  where $x_i=\varphi_{t_i}(x_0)$. Hence $x_{i+1}=F(x_i)$ for all $i$. We choose a $C^1$ real-valued test
function $\Phi$ and generate the time series $z_i=\Phi(x_i)$. (For example, if $x_i$ are points in some Euclidean space  $\reals^n$, $\Phi$ is the projection onto one of the coordinates.)
We
associate to this time series the delay coordinate vectors:
%
\[t_i\mapsto (z_i, z_{i+\tau}, z_{i+2\tau},\ldots, z_{i+(d-1)\tau}),\]
with $d$ is a sufficiently large embedding dimension and $\tau$ is a {\em lag}.  Alternately, we can think of $(z_i, z_{i+\tau}, \ldots, z_{i+(d-1)\tau})$ as a sliding window for
the time series.
On this set of delay-coordinate vectors $\mathcal{Z}$ we define
a dynamical system given by the shift map
\[s (z_i, z_{i+\tau}, \ldots, z_{i+(d-1)\tau})= (z_{i+\tau}, z_{i+2\tau}, \ldots, z_{i+d\tau}).\]
Thus, we represent the original dynamical system by a discrete dynamical system $(\mathcal{Z},s)$, whose phase space
consists of delay coordinate vectors and the mapping is the shift map above.

First, consider the limit case when $\sigma=0$ and $\varepsilon=0$
(ODE with time-independent parameter).  The Takens Embedding
Theorem~\cite{Takens1981} and its extensions by Sauer, Yorke,
Casdagli~\cite{sauer1991embedology}, imply that, for generic $\Phi$,
and for $\Delta t$ small enough, there exists a sufficiently large $d$
such that $F$ on $A_{\lambda}$, $\lambda\neq\lambda^*$, is conjugate to
$s$ on some subset of delay coordinate vectors $\mathcal{Z}$. In other
words, the shift dynamics on delay coordinate vectors is an embedded copy of the
original dynamics.

Now, if the parameter $\lambda(t)$ varies sufficiently slowly and the
noise intensity is small enough, i.e., $\varepsilon$, $\sigma$ are
sufficiently small, 
sufficiently large $t$-time interval, the dynamical system of $F$
follows closely a quasi-static attractor $A_{\lambda(t)}$, for
$\lambda(t)<\lambda^*-\delta$. The reconstructed dynamics provides an
approximate copy of $A_{\lambda}$. The same assertion holds for
$A_{\lambda(t)}$, $\lambda(t)>\lambda^*+\delta$.  Thus, the bifurcation in
the deterministic system \eqref{eq:ode-param} will be reflected in
\eqref{eq:gen-sys} by the change in the topological features of the
quasi-static, noisy attractors~\cite{ChenFu_randomDS} in the interval
$\lambda \in (\lambda^* - \Delta, \lambda^* +\Delta)$.  In what
follows, we describe the topological tools that can be used to
measure, in a robust way, the changes in the topology of these
attractors in the neighborhood of a bifurcation.

We regard each delay coordinate vector (sliding windows)  $\mathbb{X}_i=(z_i, z_{i+1}, \ldots, z_{i+d-1})$ as an element of a point cloud data set, and  we want to describe the topology of these point cloud data for all $i$. We are interested in determining from changes in the topological features of these point cloud data whether the underlying system undergoes a bifurcation of the type described above.

\subsection{From point cloud data  to persistent topology}
\label{sec:persistence}

\begin{figure}
  \centering
  \includegraphics{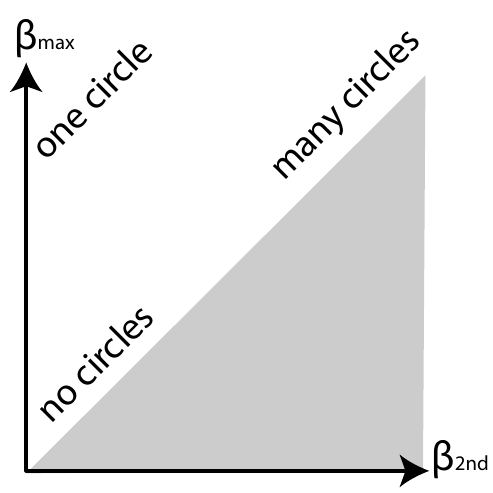}
  \caption{{\em Topological types as identified by degree 1 persistence
    lengths. For higher circle counts, more Betti numbers can be used
    for a larger space for detecting topological type.}}
  \label{fig:phasespace}
\end{figure}

To each cloud point data $\mathbb{X}_i$ as described in the previous
section, we associate an algebraic representation of its topological
features. The pipeline, shown schematically in~\eqref{eq:pipeline}, is
the following: From the point cloud data one constructs a sequence, or
{\em filtration} $\mathbb{V} =
\{VR_{\varepsilon}(\mathbb{X})\}_{\varepsilon}$, of simplicial complex
(Vietoris-Rips complex) which depends on a parameter $\varepsilon$
(which can be thought of as the `resolution level' for the data). The
simplicial homology of each complex in $\mathbb{V}$ is then
computed. The key step of persistent homology is following the
homology generators as the parameter $\varepsilon$ varies. The output
of this process is a diagram that summarizes the `birth' and `death' of homology generators; this diagram is referred to as a
\emph{barcode} or a {\em persistence diagram}. 

  \begin{align}\label{eq:pipeline}
    \mathbb{X} \ra 
    \{VR_{\varepsilon}(\mathbb{X})\}_{\varepsilon} \ra 
    \text{Persistence diagram}  
  \end{align}

We now provide some necessary background on homology to aid the reader
in understanding the information contained in a persistence diagram.
Homology is a classical technique for topological feature
identification using linear algebra. A triangulated space gives rise
to a vector space $C_d$ of \emph{chains}: formal linear combinations
of simplices. The geometric boundary gives rise to a linear boundary
operator defined as $\partial:C_k\to C_{k-1}$,
\[
\partial_k [x_0,\dots,x_k] = \sum_j(-1)^j [x_0,\dots,\hat{x_j},\dots,x_k]
\]
where by conventio $\hat{v}_j$ means leaving
the vertex $v_j$ out. The definition extends by linearity to the
entire chain space. The $k$-th \emph{homology}  quotient vector space
$H_k = \ker\partial_k/\img\partial_{k+1}$. The $k$-th Betti number is
the rank of $H_k$
$\beta_k=\textrm{rank}(H_k)$.
 The $1$-st Betti number, which is the rank of $H_1$,  counts the
number of 1-dimensional holes (`tunnels')  in $X$. Similarly $\beta_2$ counts the number of  2-dimensional holes (`cavities')  in $X$.

For a point cloud $\mathbb X$, the $\varepsilon$-Vietoris-Rips complex
$VR_\varepsilon(\mathbb X)$ is a simplicial complex with vertices
given by the points in $\mathbb X$, and a simplex $[x_0,\dots,x_d]$
included if and only if $d(x_i,x_j)<\varepsilon$ for all pairs
$x_i,x_j$ of vertices. As $\varepsilon$ grows, the Vietoris-Rips
complex gains more simplices, producing a diagram of inclusions
\[
VR_{\varepsilon_0}(\mathbb X) \hookrightarrow
VR_{\varepsilon_1}(\mathbb X) \hookrightarrow
VR_{\varepsilon_2}(\mathbb X) \hookrightarrow
VR_{\varepsilon_3}(\mathbb X) \hookrightarrow \dots
\]

\begin{figure}[ht]
  \centering
  \includegraphics{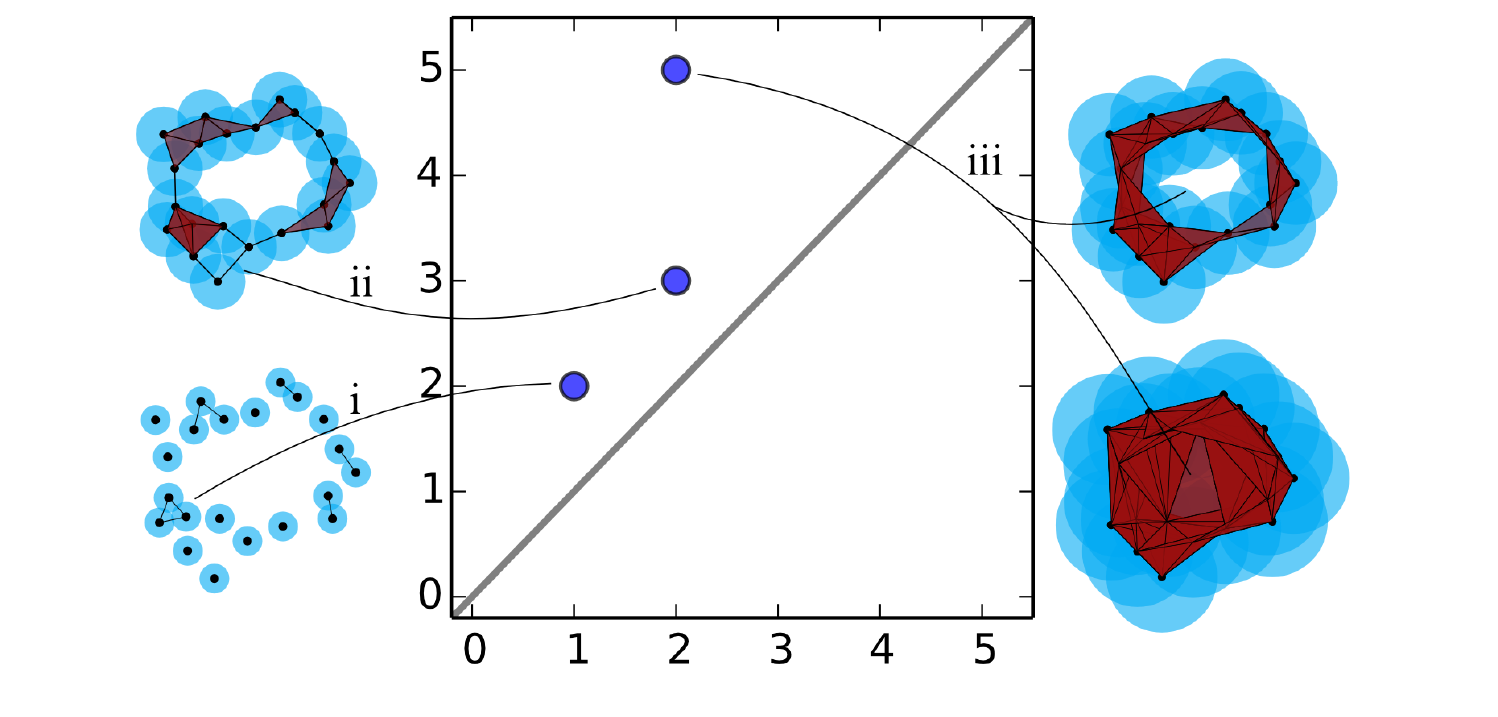}
  \caption{A sample persistence diagram for the point cloud represented by the black dots. Radii increase clockwise from lower left. See text for details.}\label{fig:rips_example}
\end{figure}

The technique of \emph{persistent homology} introduced by
\textcite{edelsbrunner_topological_2002} produces efficient algorithms
to compute the homology of a diagram of spaces like this, summarizing
it with a \emph{persistence diagram}: a multiset of start and
endpoints such that if $[\varepsilon_b,\varepsilon_d)$ is in the
  diagram, then there is a corresponding homology class that exists in
  all $VR_\varepsilon(\mathbb X)$ for $\varepsilon_b\leq \varepsilon <
  \varepsilon_d$. Long lifespans (eg., $\varepsilon_d -
  \varepsilon_b$) correspond to dominant topological features; short
  lifespans correspond to noise or small features. The coordinates $\varepsilon_b$ and
  $\varepsilon_d$ are termed {\em birth} and {\em death} times. (In
  this context, time refers solely to the monotone-increasing radius of
  the $\varepsilon$-balls used to construct the
  $\varepsilon$-Vietoris-Rips complex.)

For a classic overview of algebraic topology and homology, we recommend
the book by \textcite{hatcher_algebraic_2002}. For good overviews of
persistent homology and its use in analyzing point clouds we recommend
the survey articles by \textcite{ghrist_barcodes:_2008} or
\textcite{carlsson_topology_2009}, or the books by
\textcite{edelsbrunner_computational_2009} and
\textcite{zomorodian_topology_2005}. For a general overview of
computational topology see \textcite{Kaczynski2004}.

An example of persistent homology on a point cloud, and the corresponding persistence diagram is provided in Figure~\ref{fig:rips_example}. The size of the blue disks centered on the black vertices corresponds to $\varepsilon$: Moving clockwise, the radii increase in size, beginning in the lower left corner, which yields a nested sequence of VR complexes. At a radius of $\varepsilon=1$, a small circle is born, which subsequently dies at $\varepsilon=2$. The feature is recorded by the point at $(1,2)$, with the curve $i$ connecting the feature to its representation on the persistence diagram in the center. Another small feature exists for a single time step, indicated by line $ii$. Last, the large, central circle is also born at $\varepsilon=2$, but persists until it fills in at $\varepsilon=5$. This is shown in the two VR complexes on the right hand side and the curve $iii$. This final lifespan is significantly longer than the first two, implying that the corresponding feature is a dominant topological feature in the space.

\subsection{From persistent topology to machine learning}
\label{sec:machine-learning}

The sequence of steps described above start with a sliding window
along the time series and produce a topological summary encoded by a
barcode. As the underlying system undergoes a bifurcation, the corresponding attractors experience topological changes that are reflected by barcode diagram changes. 
We want to be able to distinguish significant changes in the barcodes that can be used as
indicators of bifurcations. For this purpose we use machine learning techniques.

Machine learning aims to reconstruct, or \dfn{learning}, a discrete
(\dfn{classification}) or continuous (\dfn{regression}) function on
some space given samples drawn from a distribution on that space. A
rich toolbox has been developed to learn functions in various
cases. In this paper we focus on using classifiers -- our goal is to learn
a discrete classification on the time series data under
study.

Classifiers for discrete data can be \dfn{linear} or \dfn{non-linear},
depending on whether they can be modeled with a linear hyperplane as a
\dfn{decision boundary} (delimiting the regions of input values that
produce different results) or not. Furthermore, learning methods can
be \dfn{unsupervised}, \dfn{semi-supervised} or \dfn{supervised}. For
a supervised problem, sample points are drawn together with their
expected values, and the system learns to generalize from seen
examples to unseen examples: one example is linear regression or
interpolation type problems. An unsupervised problem expects the
machine learning algorithm to create some set of labels on its own: a
typical example is most clustering algorithms in widespread use. For a
detailed overview of machine learning topics we refer to the excellent
textbook by Bishop \cite{christopher2006pattern}.

\section{Learning topological differences}
\label{sec:learn-topol-diff}

A periodic or quasiperiodic multi-dimensional dynamical system under
the influence of noise will tend to trace out a space with non-trivial
degree 1 homology. The easiest example is given by the simple
periodicity found in, e.g., the $(x,\dot x)$-plane of a simple
pendulum: the periodic regime traces out a simple closed curve in the
phase space of the system. A simple pendulum driven by a periodic
force of sufficiently irrational period traces out the surface of a
torus in $(x, \dot x, \ddot x)$. In these examples and in more general
cases, the presence of a non-trivial degree 1 homology group in the
point cloud traced out by a time series measurement of the system can
be correlated to the presence and type of periodicity exhibited in the
system.

\begin{figure}[hb]
  \centering
  \begin{subfigure}[b]{0.33\textwidth}
    \includegraphics[width=\textwidth]{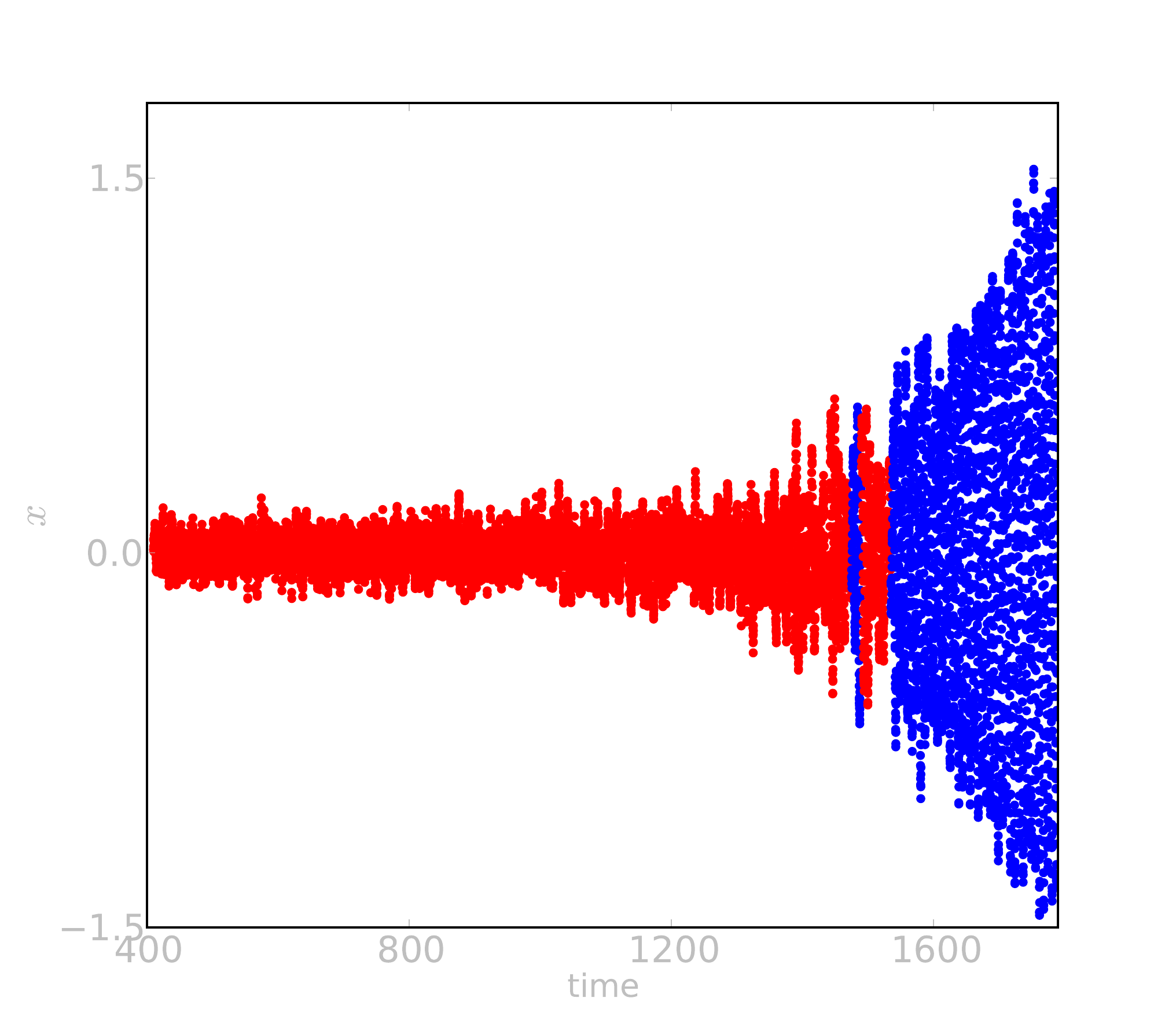}
    \caption{$k=2$}
    \label{fig:hopf2}
  \end{subfigure}%
  \begin{subfigure}[b]{0.33\textwidth}
    \includegraphics[width=\textwidth]{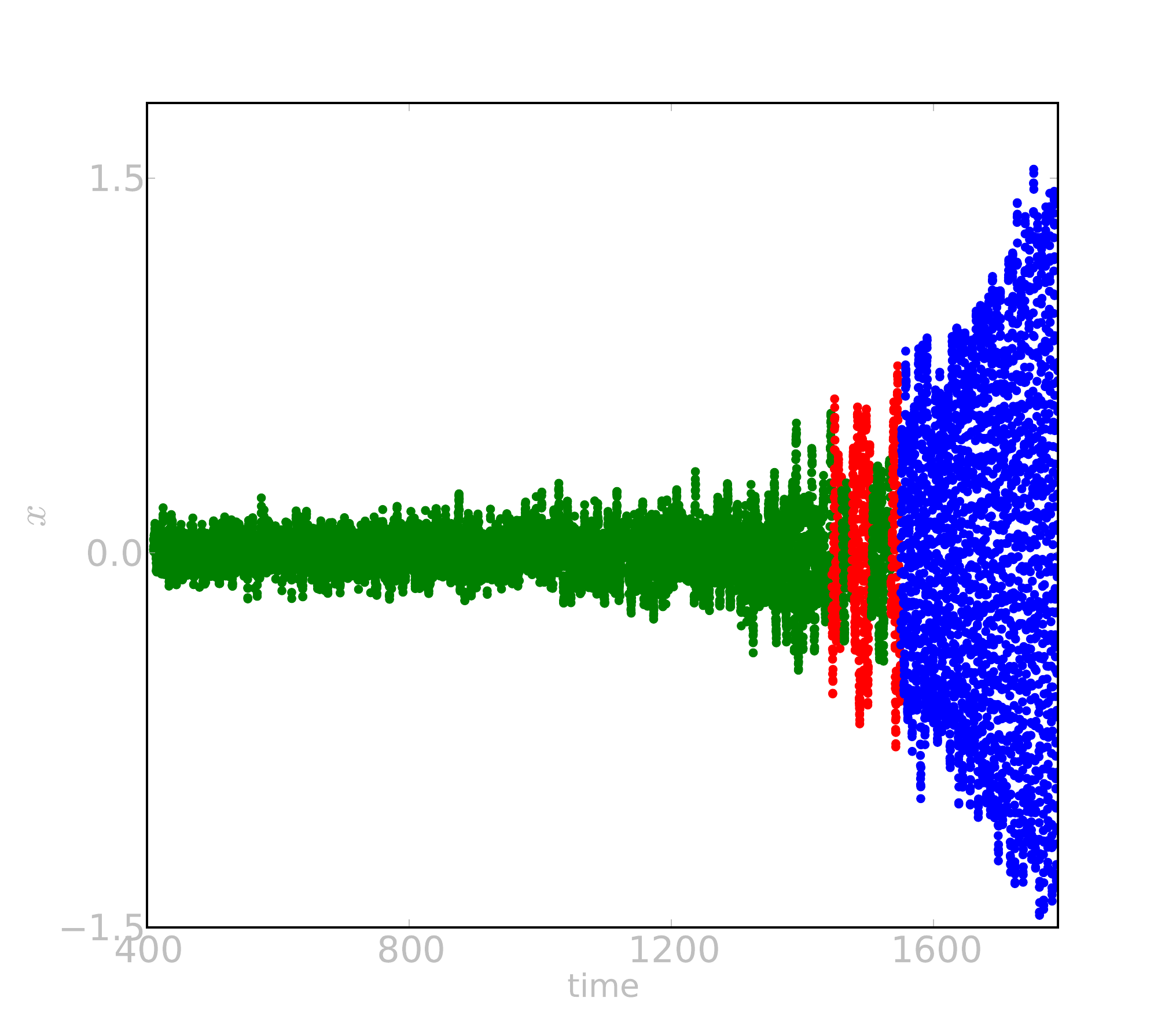}
    \caption{$k=3$}
    \label{fig:hopf3}
  \end{subfigure}%
  \begin{subfigure}[b]{0.33\textwidth}
    \includegraphics[width=\textwidth]{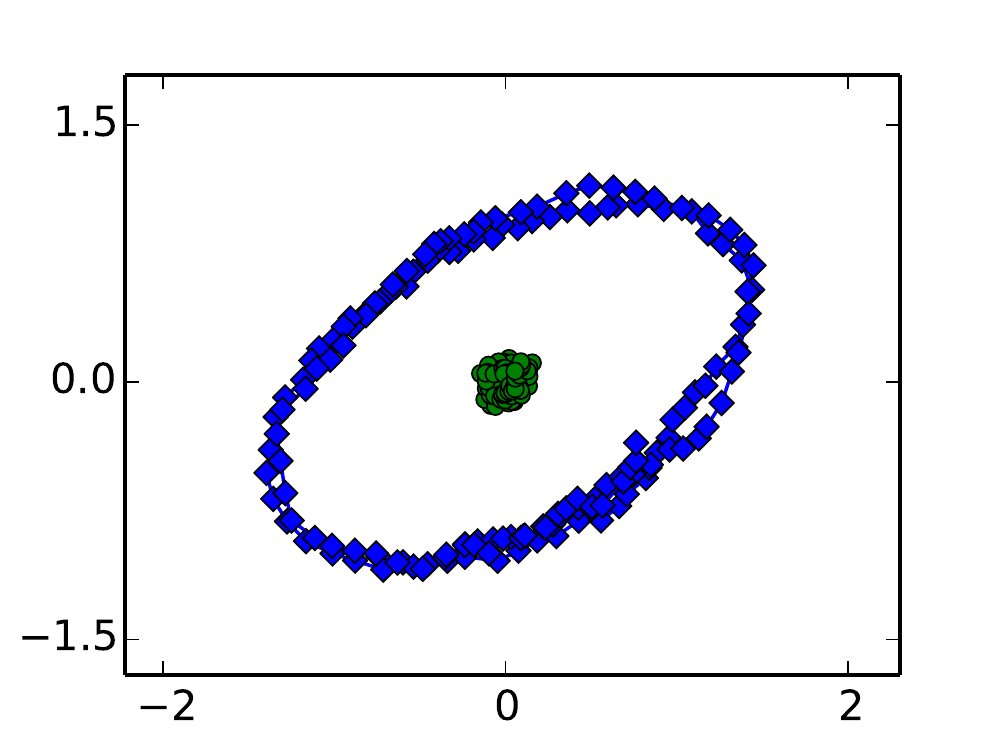}
    \caption{Before (blue circles) \newline After (green diamonds)}
    \label{fig:hopfportrait}
  \end{subfigure}%
  \caption{{\small The stochastic system in \eqref{eq:hopf} with a
      slowly varying parameter undergoes a bifurcation in which a
      stable equilibrium changes to unstable and then grows to a
      periodic orbit. We show $t$ vs. $x$. The clustering in (b)
      highlights and separates a region for which the clustering in
      (a) found questionable, mostly due to stochastic effects. In (c)
      we show a portrait of the system in the $x$-$y$--plane for one
      window taken before the bifurcation and one window taken after
      the bifurcation.}}\label{fig:hopf-tag}
\end{figure}

We aim to build a classifier capable of detecting the presence of
highly significant 1-dimensional homology classes. Examples of features
that we expect to easily discern are seen in
Figure~\ref{fig:hopfportrait}: the stationary parameter region of the
system produces no significant persistent cycles, while the periodic
regime produces a highly significant 1-cycle.  We will accomplish this
by training our classifier on the top persistence lengths of dimension
1 persistent homology. As described in the schematic in
Figure~\ref{fig:phasespace}, a high value for the most persistent
Betti number and a low value for the second most persistent Betti
number is an indication of periodic or quasi-periodic behavior, while
several high values in the top most persistent Betti numbers indicated
a more complex recurrent behavior.

For most practical computations, we pick an upper limit for the
computation -- any features that still exist when the computation
terminates are assigned a death time after the upper limit time
$t_{\max}$. Since the computation stops at $t_{\max}$, the computed
persistences will not be able to distinguish further between bars
after this point.  We choose an $r>0$ and assign all such infinite
bars a stopping value of $t_{\max}+r$, to avoid problems with infinite
values in the machine learning algorithms. (For all experiments
reported herein we chose $r=2$.) We expect to distinguish features
with very short persistence intervals from features with one or
several highly persistent features. 
Highly persistent but finite bars and infinite bars are both
indications of the presence of a significant topological feature.

To recognize periodic regimes, we use the two longest bars of the
persistence barcode as features. Intuitively, a barcode for which the
longest bar is significantly longer than the next longest bar is more
similar to a circle, thus more likely to indicate a timeseries window
from a periodic regime. In order to avoid arbitrary choices as much as
possible, we use an unsupervised learning approach on the ordered
pairs of lengths of features.

For recurrent or quasi-periodic regimes, we introduce more persistence
features, in order to distinguish between possible intermediate
regimes. For instance, in the case of the Lorenz attractor, certain
parameter values yield a two-lobed attractor. With this heuristic, we
expect to be able to distinguish (quasi-)periodicity around a single
center from the two-lobe case which is qualitatively a different
regime.

To tag a timeseries from a dynamical system with minimal operator
intervention, we can use this persistence-based feature collection as
the basis for a linear unsupervised learning system. There is a wealth
of machine learning schemes to choose from -- for simplicity, we
work with $k$-means clustering. Expecting few  dominant
 features, we train classifiers on the top 2 or top 3 most persistent bar lengths.
Our rationale for this was that assigning a
hypothetical label of ``high persistence'', ``medium persistence'' or ``low persistence'' to
each of the top three values would produce exactly 10 possible ordered
sequences of 3 labels. In practice, the tagging regimes tend to
stabilize for our examples above 3 labels.

We imagine that if the analysis calls for it, a different unsupervised
or semi-supervised method may well be used; semi-supervised methods
requiring more effort to tag the supervision part of data input. For
our test cases, however, already $k$-means performed well beyond our
expectations -- see Section~\ref{sec:results} for details -- and
we save the exploration of available machine learning methods for future work.

An important detail that we note is that we fully expect these methods
to break down once the periodicity length significantly exceeds the
window size.

\begin{figure}[h]
  \centering
  \begin{subfigure}[b]{0.33\textwidth}
    \includegraphics[width=\textwidth]{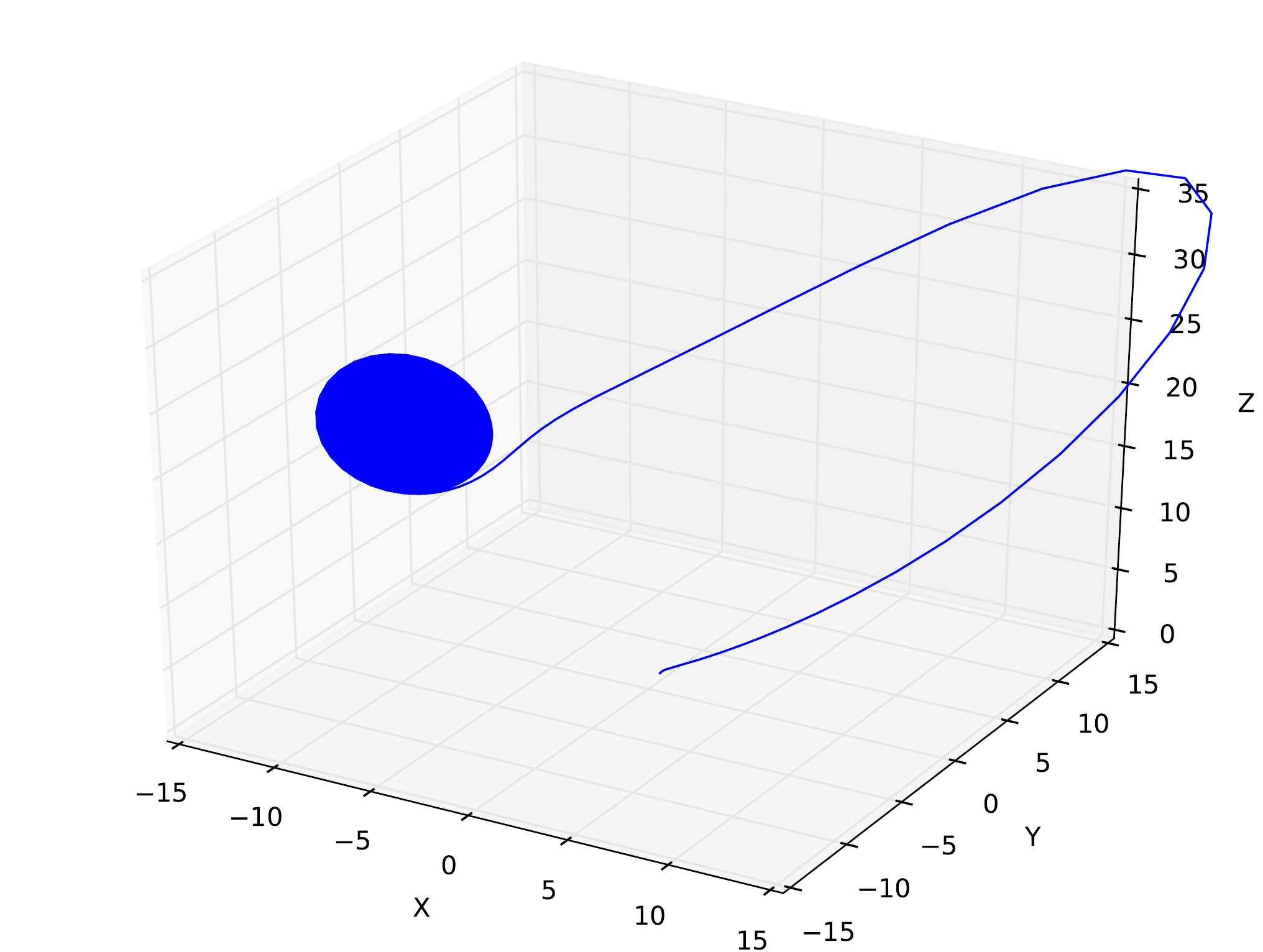}
    \caption{$\rho=23.5$}
    \label{fig:lorenz235}
  \end{subfigure}%
  \begin{subfigure}[b]{0.33\textwidth}
    \includegraphics[width=\textwidth]{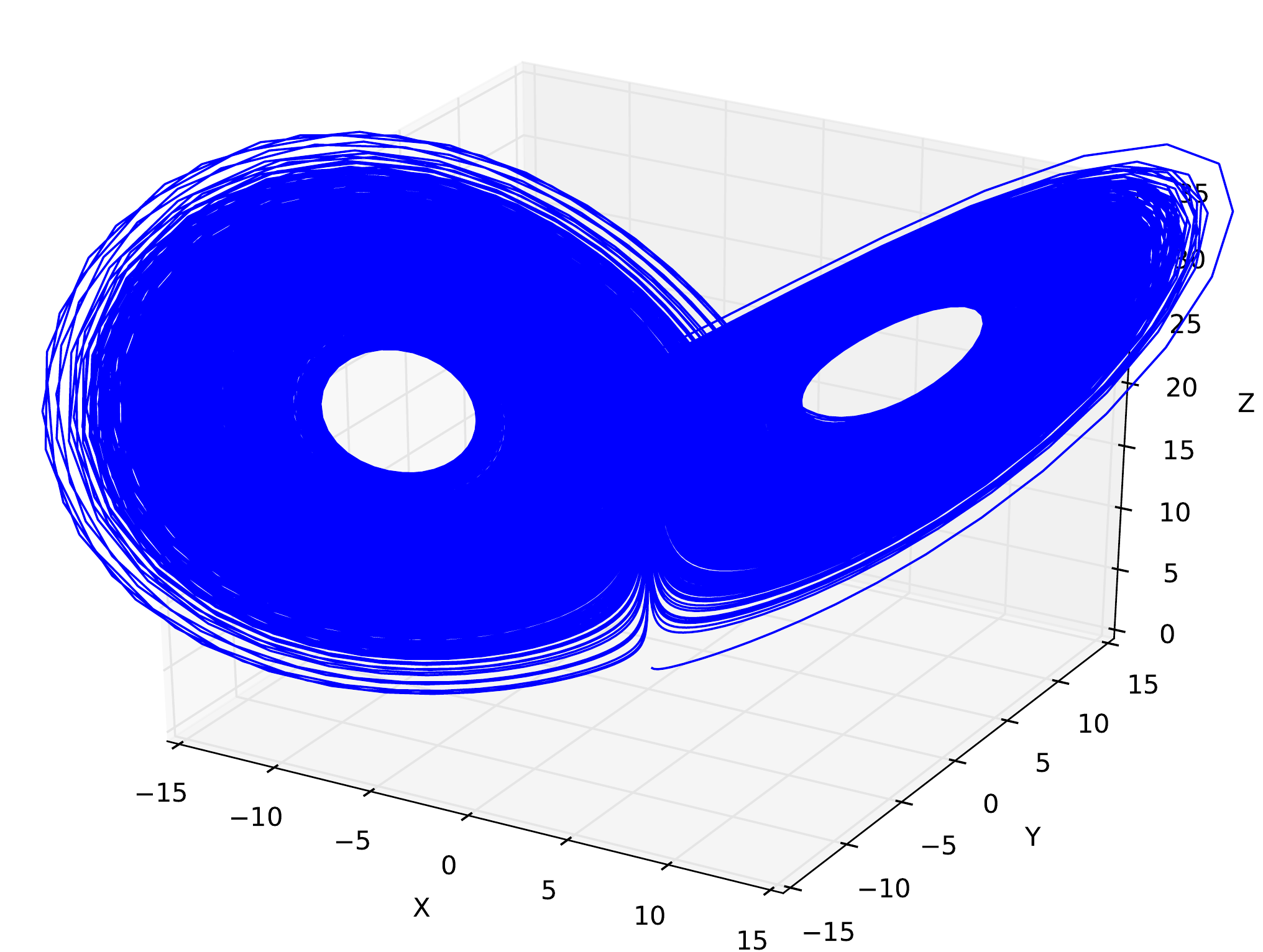}
    \caption{$\rho=24.5$}
    \label{fig:lorenz245}
  \end{subfigure}%
  \caption{{\small The Lorenz system undergoes a global bifurcation as $\rho$
    increases from 23.5 (a) to 24.5 (b).}}
  \label{fig:lorenz}
\end{figure}

\section{Results}
\label{sec:results}

We demonstrate the effectiveness of the persistence-based automatic
tagging algorithm on three experiments. First, we analyze a
non-chaotic, stochastic system $h$, which undergoes a local Hopf
bifurcation. Second, we explore the detection of global bifurcations
in a chaotic system, namely the Lorenz attractor. We vary the $\rho$
parameter through a global bifurcation, from $\rho=23.5$ to
$\rho=24.5$~\cite{Doedel2006}. Lastly, we investigate temperature and
CO$_2$ records obtained from the Vostok ice core~\cite{vostok}

\subsection{Hopf bifurcation}
\label{sec:hopf}

There are many systems that exhibit a Hopf bifurcation. Without loss
of generality, we may consider the following stochastic system,
\begin{align}\label{eq:hopf}
  dx &= f(x,y) + \sigma_1 dW_1\\
  dy &= g(x,y) + \sigma_2 dW_2,
\end{align}
where $\sigma_1,\sigma_2$ represent noise level and $W_1$ and $W_2$
are one-dimensional Weiner processes~\cite{Higham2001}.  This system,
with $f(x,y) = \lambda(t)x-y-xy^2$ and $g(x,y) = x+\lambda(t)y - y^3$ and
$\lambda$ varying linearly with time $\dot\lambda=\varepsilon$, for $\varepsilon $ small, is a classic model of biological oscillators. A realization is plotted in Figure~\ref{fig:hopf-tag}.
We note that the corresponding deterministic system ($\sigma_1=\sigma_2=0$) with time-independent parameter $\lambda$ ($\varepsilon=0$)  undergoes a Hopf bifurcation for $\lambda=0$.


As we vary the parameter $\lambda$, the system progresses from a noisy
but stationary regime to a periodic regime, tracing out widening
circles in the $x$-$y$--plane.  In~\cite{BerwaldGidea2013}, Berwald
and Gidea use these topological changes in conjunction with distance
metrics on persistent homology bar codes on similar point cloud
windows to detect changes in the trajectory as $\lambda$ drifts in
time. In the current manuscript, we take take this a step further by
applying the unsupervised learning tools described above to cluster
the data windows in the time series. We illustrate the results for
example windows from the two regimes in Figure~\ref{fig:hopf-tag}. In
Figure~\ref{fig:hopf2} we use 2-means clustering, and the learning
algorithm is confident in locating the bifurcation in terms of
the growth of the orbit, only faltering when stochastic effects cause
the cycle radius to decrease for a short period of time around
$t=1500$. In Figure~\ref{fig:hopf3}, by using $k=3$ the same region
that caused an issue in Figure~\ref{fig:hopf2} is highlighted by the
intermediate cluster. One of the strengths of our method is its
ability to highlight regions of uncertainly in the data. In this case,
the level of noise contributes to uncertainty as the bifurcation grows
into a limit cycle, which is exactly the region that we would like
our algorithms to locate.

\subsection{Lorenz equations}
\label{sec:lorenz}

\begin{figure}[ht]
  \centering
  \begin{subfigure}[b]{0.33\textwidth}
    \includegraphics[width=\textwidth]{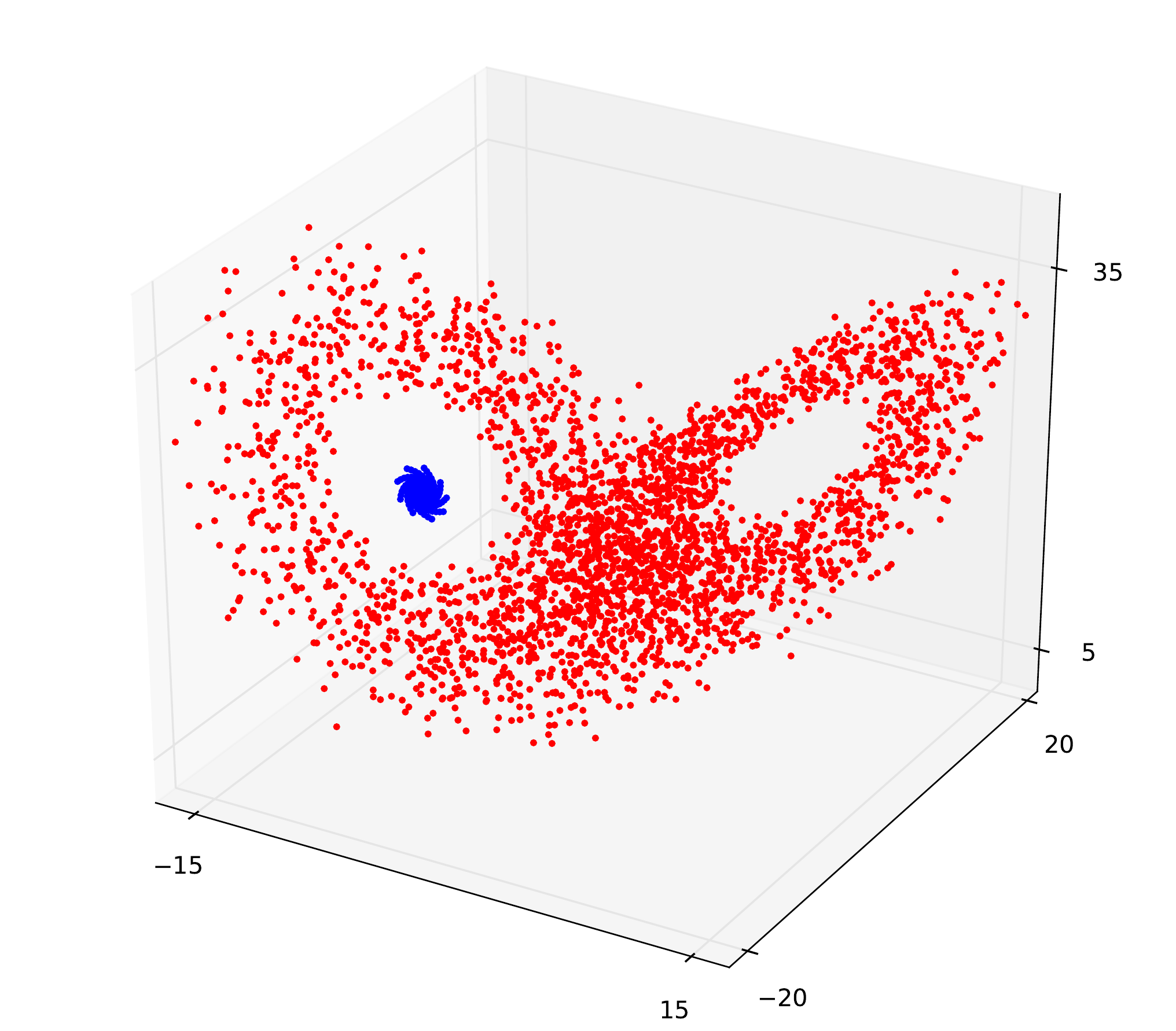}
    \caption{$k=2$}
    \label{fig:lorenz2}
  \end{subfigure}%
  \begin{subfigure}[b]{0.33\textwidth}
    \includegraphics[width=\textwidth]{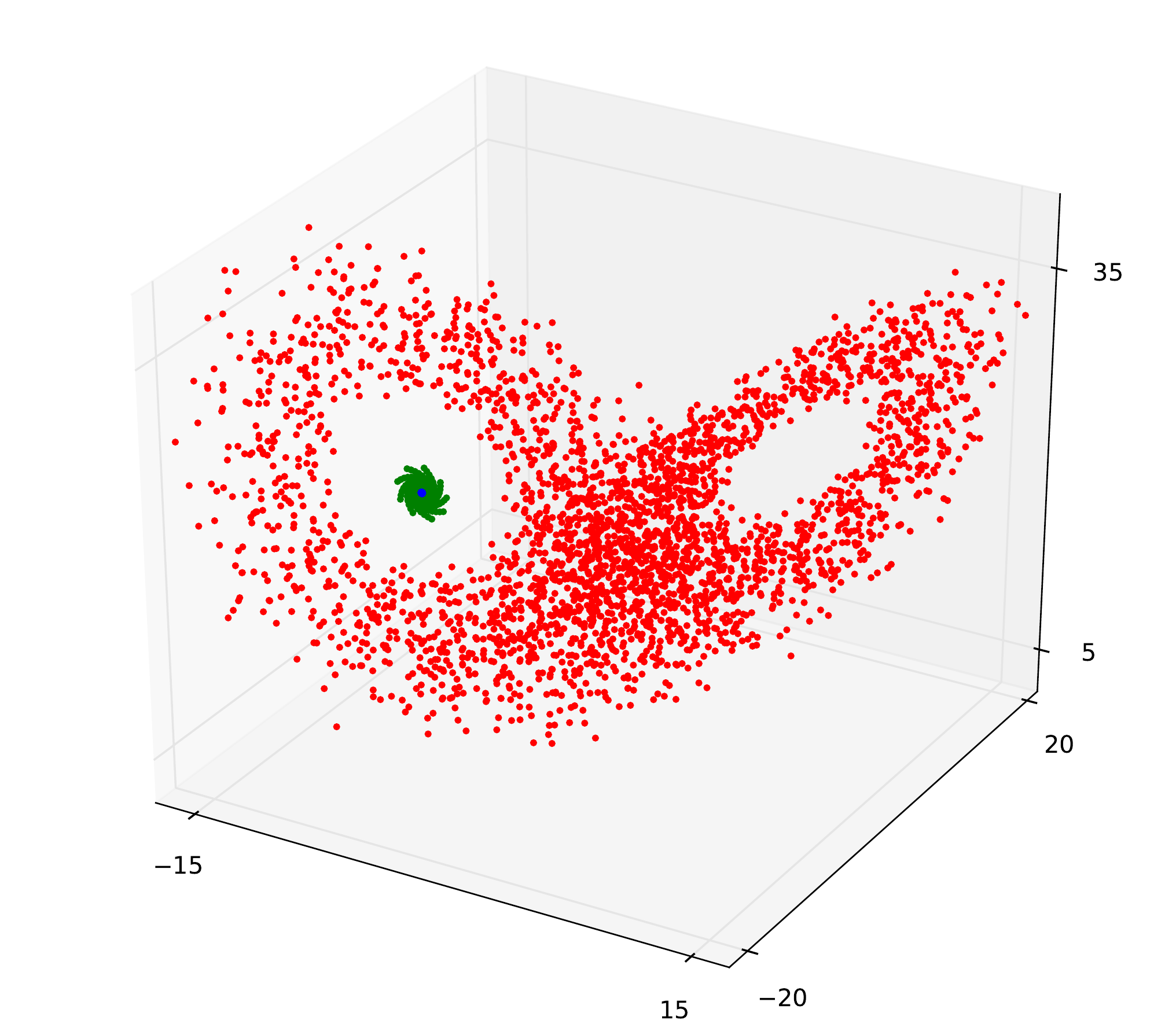}
    \caption{$k \ge 3$}
    \label{fig:lorenz3}
  \end{subfigure}%
  \begin{subfigure}[b]{0.33\textwidth}
    \includegraphics[width=\textwidth]{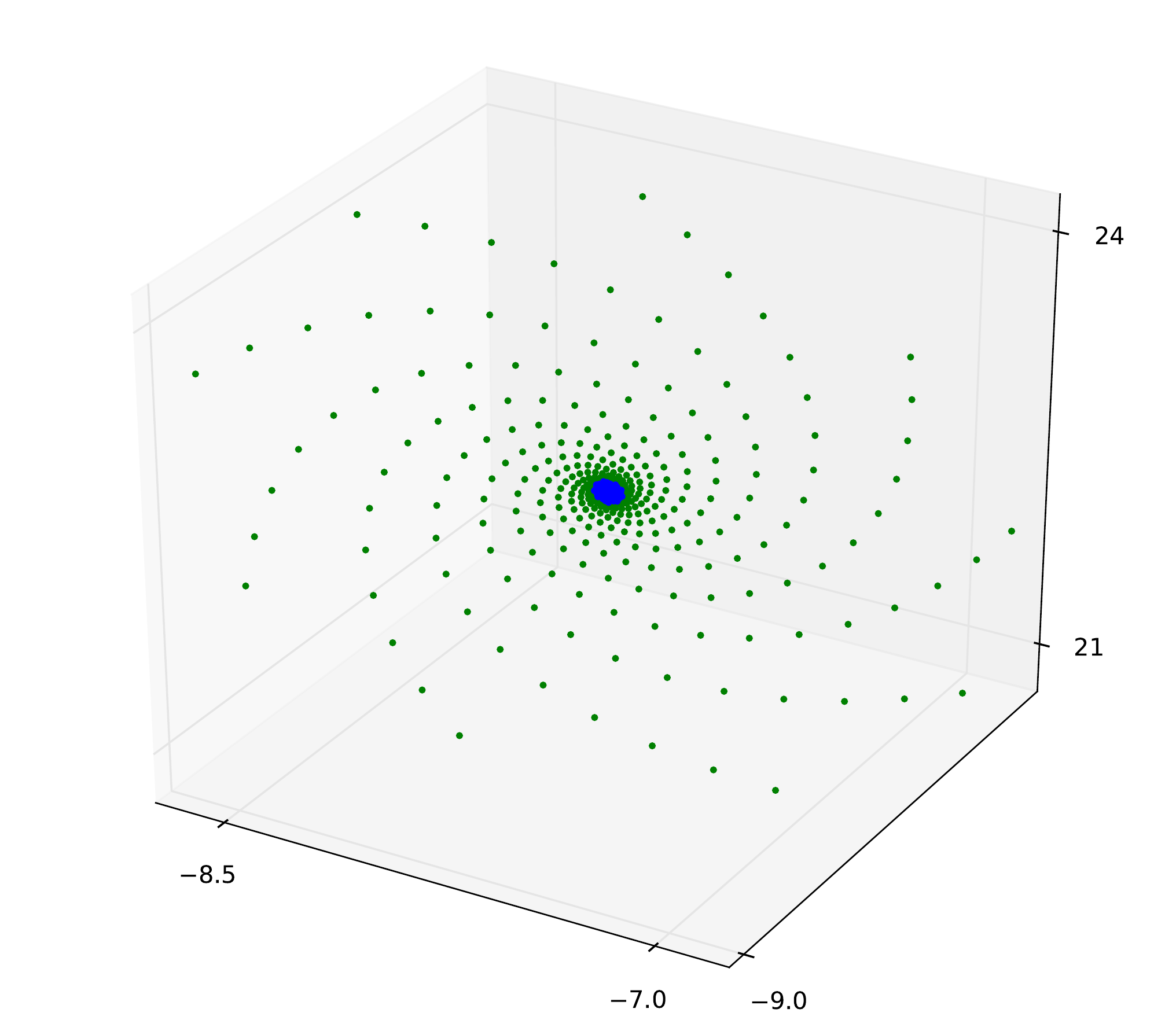}
    \caption{Detail for $k \ge 3$}
    \label{fig:lorenz3_zoom}
  \end{subfigure}%
  \caption{{\small Using a $2$-means classifier in (a) we are able to
      distinguish the stationary regime at 23.5 (blue) from the
      recurrent regime at 24.5 (red). With a classifier using at least
      3 means, in (b), we get exactly 3 classified regimes: the
      stationary core of the regime at 23.5 (blue), the slow spiral
      into this stationary core (green) and the two lobes of the
      recurrent regime at 24.5 (red). (Note: we remove a transient
      from the trajectory.)}}\label{fig:lorenz_tag}
\end{figure}

The Lorenz equations have a long history of study in dynamical systems
and provide a fruitful test bed. The equations were derived by Edward Lorenz in the 1960s and represent a simplified model of atmospheric convection. Serendipitously, Lorenz discovered that they exhibit sensitivity to initial conditions, a finding which launched decades of research into chaotic attractors. The equations are defined by:
\begin{align}\label{eq:lorenz}
  \dot{x} &= \sigma(y-x)\\
  \dot{y} &= x(\rho -z)-y\\
  \dot{z} &= xy - \beta z
\end{align}
where $\sigma,\beta,\rho$ are real-valued parameters. We fix $\sigma=10$ and $\beta=8/3$, their ``classic'' values. We change $\rho$ so as to reorganize the unstable manifold. In particular, when $\rho=23.5$ we observe a trajectory that approaches one of two stable fixed points for initial conditions different from (0,0,0). Alternatively, when $\rho=24.5$, the trajectory organizes itself on a stable manifold that resembles closely the classic ``butterfly wings'' of the Lorenz attractor at the classic value of $\rho=28$. (The choice of $\rho=24.5$ was inconsequential to the topological conclusions of the tagging algorithms -- we could have very well chosen $\rho=28$.)

A trajectory for $\rho=23.5$ is shown in
Figure~\ref{fig:lorenz235}. The trajectory and asymptotic behavior
depend on the initial condition. The dependence exhibits an important
and well-known symmetry, manifesting as a rotation by $\pi$ about the $z$ axis,
\[
(x,y,z) \mapsto (-x,-y,z).
\]
There is another attractor, symmetric to the one in
Figure~\ref{fig:lorenz235}, and w.l.o.g. we can focus on just one of
the two attractors in our experiments. For
$\rho=24.5$, the situation is the same, except that in this case, the
attractor consists of the two lobes plotted in
Figure~\ref{fig:lorenz245}.

\begin{figure}
  \centering
  \includegraphics[width=10cm]{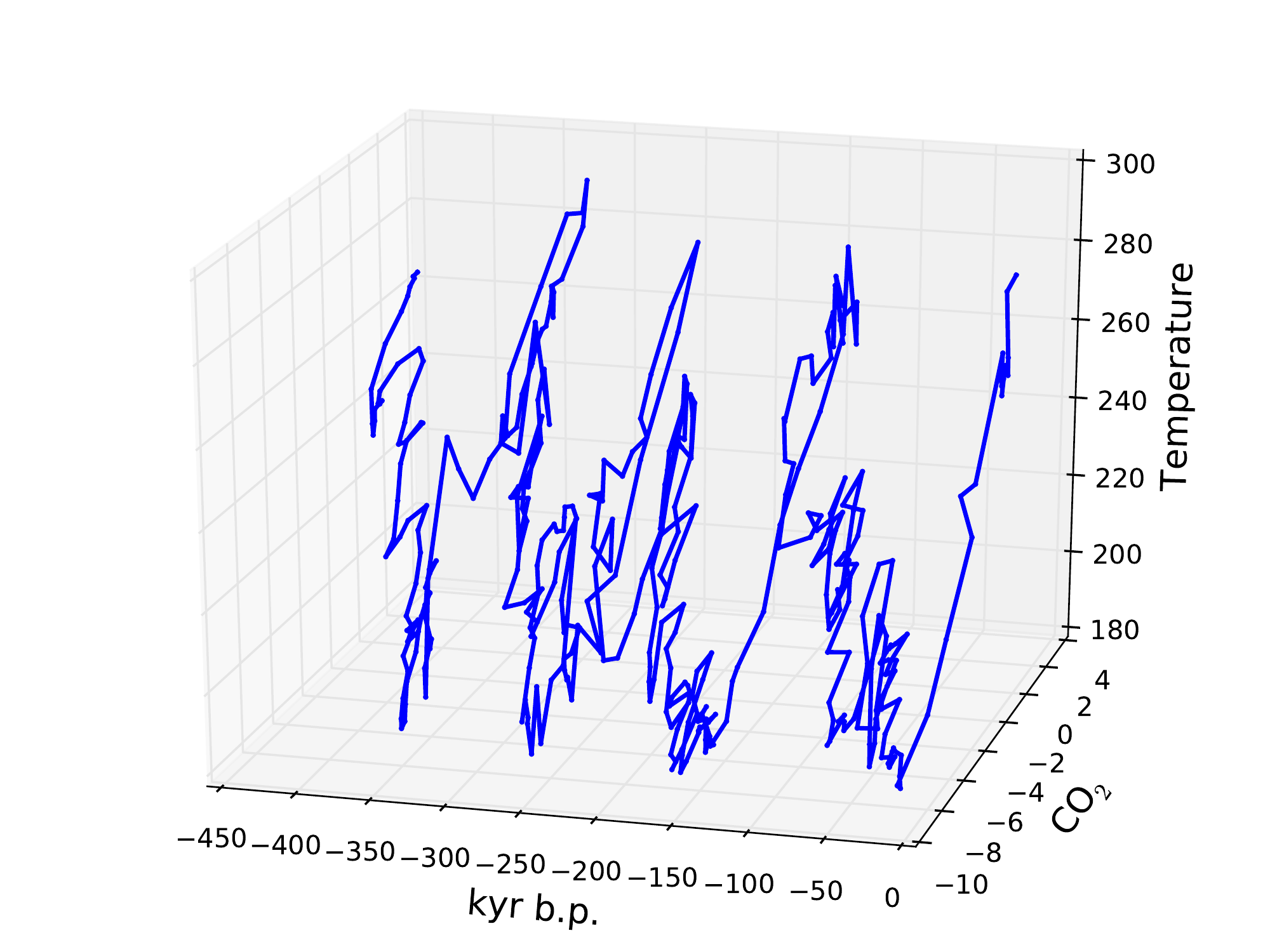}
  \caption{{\small The record of temperature and CO$_2$ over time shows
    distinct regimes, with sudden changes in both visible in both
    measurements.}}
  \label{fig:vostok}
  \vspace{-10pt}
\end{figure}
It is this global difference which is clear from a topological
perspective as well. After first removing a
 transient from the trajectory we apply the heuristics detailed in
Section~\ref{sec:learn-topol-diff} to distinguish the two parameter
regimes in an unsupervised manner. The phase space is
three-dimensional and we take as input point clouds composed of windows of
time series. We train a single unsupervised classifier
on windows from both values of $\rho$, and use the trained classifier
to tag windows from the two time series.

The result is a clean separation of the data, as shown in
Figure~\ref{fig:lorenz_tag}. For instance, with $k=2$, the data is
partitioned into two classes according to the value of
$\rho$. Increasing $k$ does little to improve the partitioning, while
it does identify the central region near the fixed point as separate
from the earlier part of the trajectory. In
Figure~\ref{fig:lorenz3_zoom} we highlight the two classes identified
for central spiral, $\rho=23.5$, seen in Figure~\ref{fig:lorenz3}. One
way to interpret these results is that quasiperiodic behavior
occurring below a certain spatial resolution is singled out as a third
class in this case.

\subsection{Vostok temperature and \ce{CO2} data}
\label{sec:vostok}

Ice cores offers a unique window into past climates on Earth. One of
the longest obtained is from the Vostok research station in
Antarctica. From this and other ice cores it is possible to
reconstruct various aspects of Earth's climate and atmosphere over
400\,000 years into the past. Two measurements possible to obtain from
an ice core are atmospheric \ce{CO2} concentration and atmospheric
temperature (often through a proxy such as an oxygen isotope ratio).
\begin{figure}
  \centering
  \includegraphics[width=\textwidth]{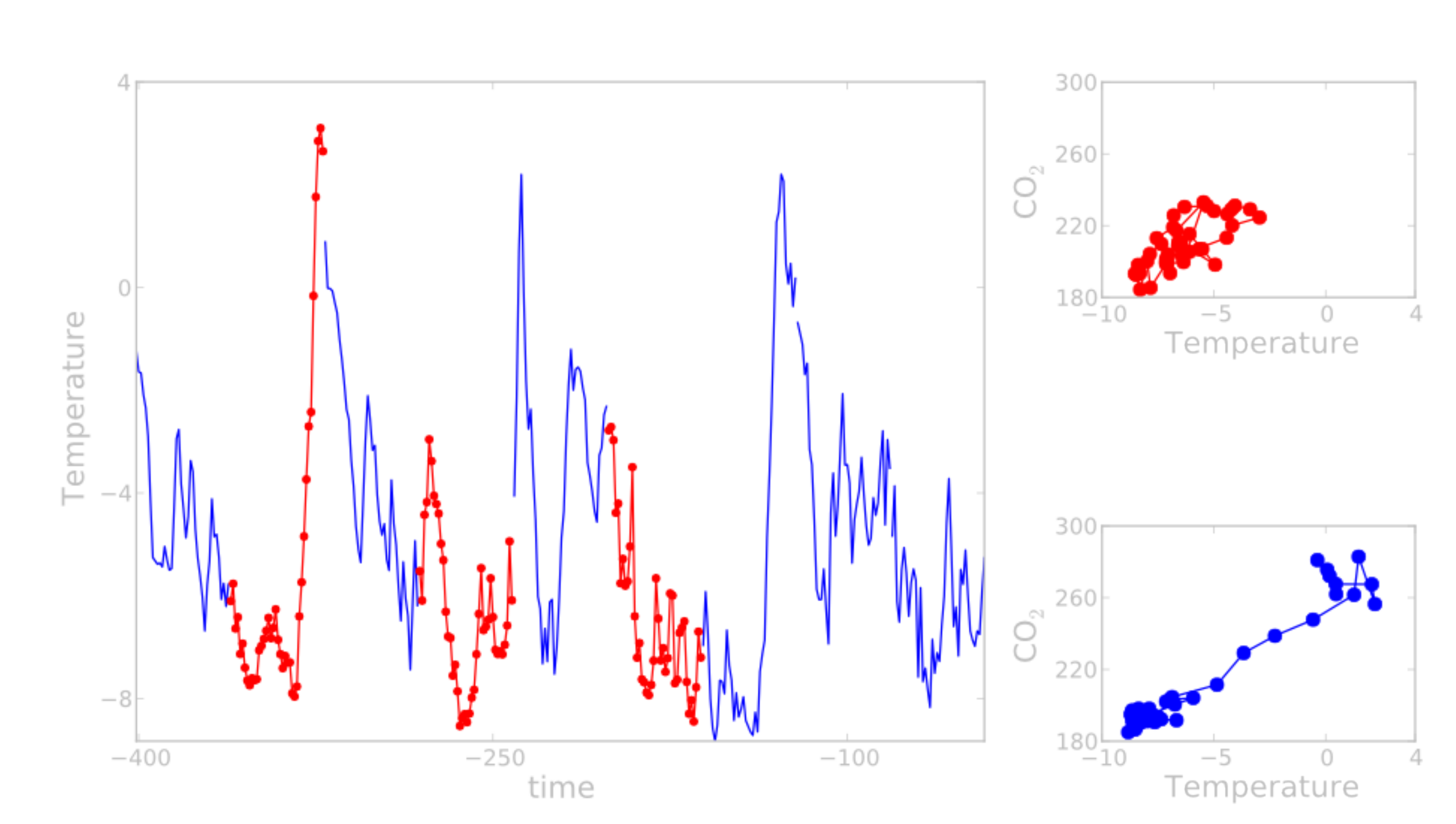}
  \caption{{\small Plot of time vs. temperature, clustered by
      persistence of cycles in the temperature -- \ce{CO2} plane (insets).}}
  \label{fig:vostok2_overlay}
\end{figure}
In the case of \ce{CO2} and
temperature, there is a lag in which rising temperatures actually
precede a rise in \ce{CO2}. When plotted in three dimensions, these lags
are observable as small-radius spiral in the time series, as seen in
Figure~\ref{fig:vostok}. These lags are a poorly understood aspect of
climate change in the geologic record.

\begin{figure}
  \centering
  \includegraphics[width=\textwidth]{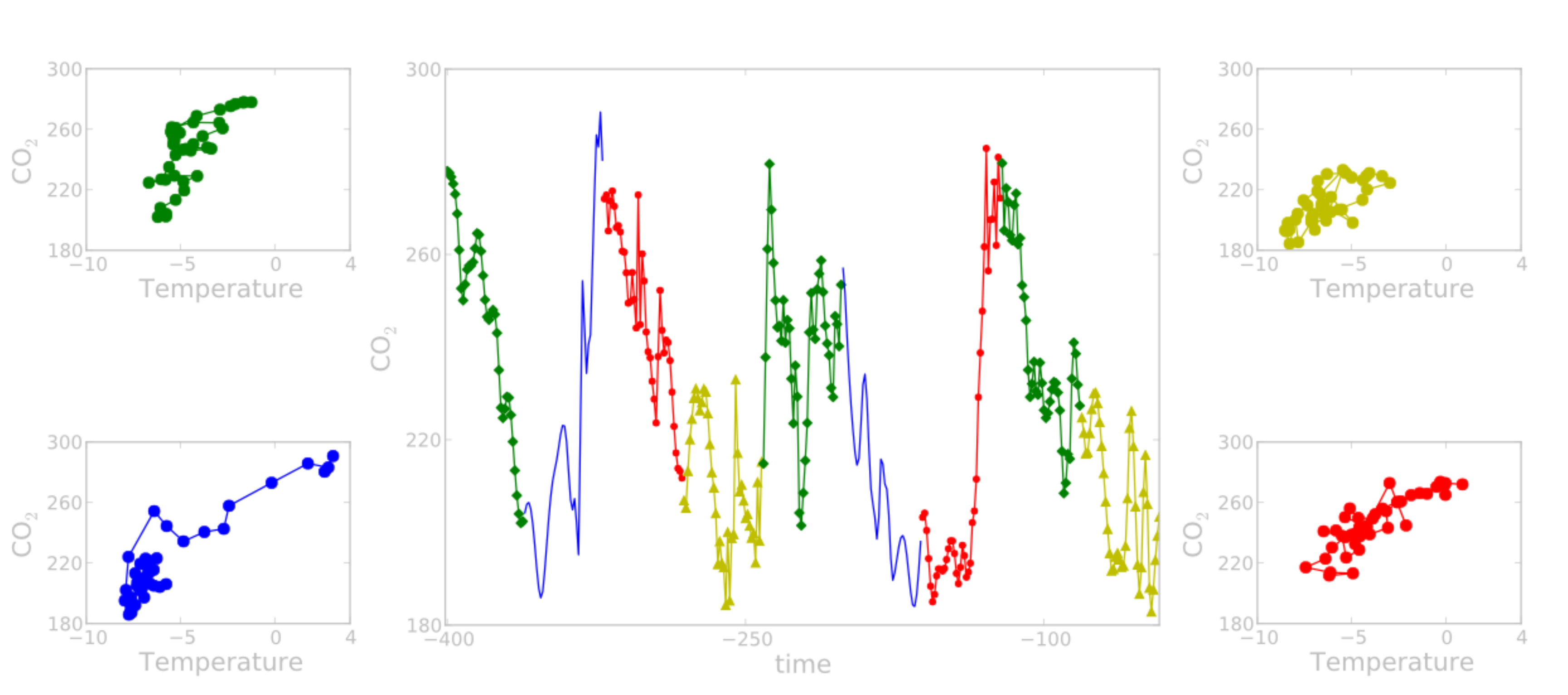}
  \caption{{\small Plot of time vs. \ce{CO2} clustered by persistence of
      cycles in the temperature vs. \ce{CO2} plane. The insets depict
      representatives of the different clusters we discovered in the
      data, plotted in their local temperature vs. \ce{CO2} coordinates.}}
  \label{fig:vostok4_overlay}
\end{figure}

By analyzing the temperature -- \ce{CO2} data windows of length in
two-dimensions, we can distinguish regimes in the climate record using
our algorithms. Combining the time record with the tagging of windows
yields the classifications shown in Figures~\ref{fig:vostok2_overlay}
and~\ref{fig:vostok4_overlay}. For values of $k \ge 3$ it is possible
to distinguish regimes similar to those found in Dakos,
\etal~\cite{Dakos2008}. In particular, we find the added granularity
of a higher number of clusters important to separate the marginally
more stationary regimes, yellow $\triangle$'s, which correspond to
regions of ``critical slowing down'' in Dakos, \etal, from regimes
that possess small loops but also a definite linear trend. Finer
granularity in the data, which would enable a shorter window size,
would likely aid in the analysis in this case. As in other work,
especially~\cite{Dakos2008}, the sparsity of the data can be a
hindrance to exacting analysis.

The value of an analysis like this is similar to the analysis
exhibited for breast cancer in~\cite{lum_extracting_2013}: we are able
to recognize a known distinct regime in the Vostok data set -- and
also several other regimes that recur in the data set with internal
homogeneity. The topological and machine learning based perspective
emerges as a way to highlight known, as well as possible new, patterns
in the data.

\section{Conclusion}
\label{sec:conclusion}

We have demonstrated that combining persistence barcode lengths as
features with an unsupervised machine learning protocol produces
strong results in dynamical systems models exhibiting both global and
local bifurcations and in automatically recognizing and tagging
different regimes for time series from dynamical systems. The
topological feature selection is robust to noise and gives a powerful
predictor for bifurcation values in noisy system when many
realizations may be computationally difficult.  In
Section~\ref{sec:hopf}, by tagging distinct regimes based on recurrent
behavior, the classification scheme is able to handle the uncertainty
of introduced by the noise by assigning the majority of that region a
separate class.

We show also that global bifurcations, from stable to chaotic regimes
for instance, can be detected in an unsupervised manner in
Section~\ref{sec:lorenz}. In this case we are concerned with the
topological changes in the stable manifold, which are large enough so
that we focus solely on the deterministic system. We showed that the
classification for such a bifurcation is extremely precise, with no
overlap in the two regimes. Extending results from
Section~\ref{sec:hopf} to a more general case, stochastic version of
the Lorenz equations is part of ongoing research.

Furthermore, the approach using computational topology as feature
selection and machine learning techniques for unsupervised
classification has proven to produce interesting results on real world
data sets. We showed that learning algorithms are able to distinguish
regimes have previously been distinguished statistically by Dakos,
\etal. In addition, by choosing different numbers of clusters, we are
able to partition the data based on topological similarity. While this
does not directly answer certain geological questions concerning
temperature or \ce{CO2} in this case, identifying new and distinct
regimes is important its own right.

\section*{Acknowledgements}

JJB would like to thank Dr. Richard McGehee for providing the Vostok ice core data.

\printbibliography

\end{document}